\documentclass[lettersize,journal]{IEEEtran}
\usepackage{amsmath,amsfonts}
\usepackage{algorithmic}
\usepackage{algorithm}
\usepackage{array}
\usepackage[caption=false,font=normalsize,labelfont=sf,textfont=sf]{subfig}
\usepackage{textcomp}
\usepackage{stfloats}
\usepackage{url}
\usepackage{verbatim}
\usepackage{graphicx}
\usepackage{cite}
\usepackage{multirow}
\usepackage{threeparttable}
\usepackage{booktabs}
\usepackage{float}
\usepackage{subfig} 
\usepackage[normalem]{ulem}
\useunder{\uline}{\ul}{}
\usepackage{longtable}\usepackage{array}
\usepackage{stackengine}

\usepackage{bbding}
\usepackage{braket}
\usepackage{qcircuit}
\usepackage{makecell}  
\usepackage{xcolor}  
\usepackage{amsthm,amsmath,amssymb}
\usepackage{mathrsfs}
\usepackage{bbding}
\usepackage{utfsym}
\begin{document}

\title{Personalized Quantum Federated Learning for Privacy Image Classification}

\author{Jinjing Shi,~\IEEEmembership{Member,~IEEE,}
		Tian Chen,~\IEEEmembership{Student Member,~IEEE,}
		Shichao Zhang,~\IEEEmembership{Senior Member,~IEEE,}
		and Xuelong Li,~\IEEEmembership{Fellow, IEEE}
		\thanks{Jinjing Shi, Tian Chen and Shichao Zhang are with the School of Electronic Information, Central South Univerisity, Changsha 410083, China (e-mail: shijinjing@csu.edu.cn; tianchen@csu.edu.cn).}
		\thanks{Shichao Zhang is also with Key Lab of MIMS, College of Computer Science and Technology, Guangxi Normal University, Guilin, 541004, China (e-mail: zhangsc@mailbox.gxnu.edu.cn).}
		\thanks{Xuelong Li is with the Institute of Artificial Intelligence (Tele AI), China Telecom Corp Ltd, 31 Jinrong Street, Beijing 100033, P. R. China (e-mail: li@nwpu.edu.cn).}}


\markboth{Journal of \LaTeX\ Class Files,~Vol.~14, No.~8, August~2021}%
{Shell \MakeLowercase{\textit{et al.}}: A Sample Article Using IEEEtran.cls for IEEE Journals}


\maketitle

\begin{abstract}
	Quantum federated learning has brought about the improvement of privacy image classification, while the lack of personality of the client model may contribute to the suboptimal of quantum federated learning. A personalized quantum federated learning algorithm for privacy image classification is proposed to enhance the personality of the client model in the case of an imbalanced distribution of images.
	First, a personalized quantum federated learning model is constructed, in which a personalized layer is set for the client model to maintain the personalized parameters.
	Second, a personalized quantum federated learning algorithm is introduced to secure the information exchanged between the client and server. 		
	Third, the personalized federated learning is applied to image classification on the FashionMNIST dataset, and the experimental results indicate that the personalized quantum federated learning algorithm can obtain global and local models with excellent performance, even in situations where local training samples are imbalanced. The server's accuracy is 100\% with 8 clients and a distribution parameter of 100, outperforming the non-personalized model by 7\%. The average client accuracy is 2.9\% higher than that of the non-personalized model with 2 clients and a distribution parameter of 1. Compared to previous quantum federated learning algorithms, the proposed personalized quantum federated learning algorithm eliminates the need for additional local training while safeguarding both model and data privacy. 		
	It may facilitate broader adoption and application of quantum technologies, and pave the way for more secure, scalable, and efficient quantum distribute machine learning solutions.
\end{abstract}

\begin{IEEEkeywords}
	Machine learning, quantum machine learning, quantum federated learning, personalized federated learning, image classification.
\end{IEEEkeywords}

\section{Introduction}
    \IEEEPARstart{F}{ederated} Learning (FL) \cite{Jakub2016}, an innovative branch of distributed machine learning, has been widely used in edge computing \cite{10316599}, image recognition \cite{10330766}, and medical diagnosis \cite{10697408}, etc. 
    Since FL trains on distributed data, it can alleviate the problem of data island (data is isolated and disconnected within an organization or enterprise) \cite{9777682}. In the training process of FL, instead of directly transmitting the original data, the gradient or model parameter information is transmitted, which can protect data privacy to a certain extent. 
    However, FL still faces two main problems. On the one hand, due to the heterogeneity of decentralized data, such as label distribution and quantity skew, the accuracy of the federated learning model may be seriously degraded \cite{10295990}. On the other hand, many attacks on gradients and model parameters make the transmission of information no longer secure, and the privacy of the original data would be threatened. For example, gradient attack can use gradient inversion attack to obtain the original data of the users, and by modifying the model parameters, the attacker would expose more local private data during the training of federated learning \cite{10274102}.
	
	Fortunately, the parallelism inherent in quantum computing and the confidentiality offered by quantum communication have introduced new developmental opportunities for federated learning, leading to the emergence of Quantum Federated Learning (QFL) \cite{Li2021}, which has become a significant research field by merging the benefits of Quantum Machine Learning (QML) with the distributed machine learning approach.
	QFL using quantum circuits or quantum channels, further develops classical FL and brings potential quantum advantages. On the one hand, QML uses the entanglement and superposition characteristics of quantum states to present the parallelism of quantum computing \cite{10613453, GQHAN}, which has the effect of exponential acceleration compared to classical computing \cite{huang2021}. The parameterized quantum circuit with only a few parameters can achieve the effect of classical machine learning models using massive parameters. On the other hand, quantum Internet has the potential to enable secure and private communication \cite{10477889}. For example, the non-cloning theorem \cite{Wootters1982} prohibits the replication of an arbitrary unknown quantum state, which helps prevent an attacker from duplicating the quantum state during communication.
	
	Due to its high efficiency and security, QFL is suitable for privacy image classification, which is an important task in distributed machine learning. However, there are two problems with privacy image classification based on QFL. One is to strike a tradeoff between the security of privacy images and the accuracy of classification \cite{9915468}, and another is to reduce the negative impact of the non-independent identically distribution (non-IID) of privacy images \cite{9743558}. In recent years, numerous studies have concentrated on how quantum federated learning can improve the accuracy and security of privacy image classification.
	In 2021, Li \textit{et al.} \cite{Li2021} proposed a QFL through blind quantum computing, where differential privacy is used to preserve the training gradient of the privacy images. In 2022, Huang \textit{et al.} \cite{Huang2022} introduced a QFL framework built on the variational quantum algorithm, which achieves high accuracy on the MNIST dataset. In 2023, Song \textit{et al.} \cite{song2023} proposed a QFL framework for classical clients using the technique of shadow tomography, thereby removing the necessity for clients to use scarce quantum computing resources in image classification. In 2024, Qu \textit{et al.} \cite{10572363} introduced a quantum fuzzy federated learning, manifesting in faster training efficiency, higher accuracy, and enhanced security. Luca \textit{et al.} \cite{Luca2024} proposed a hybrid quantum federated learning for hepatic steatosis image diagnosis. 
	However, the above studies have ignored the lack of personality of the client model in non-IID scenarios, which leads to the suboptimal performance of QFL.
	
	To enhance client model personalization while maintaining the security of private images in FL, we propose a Personalized Quantum Federated Learning (PQFL) model. By introducing a personalized layer, the client model retains more personalized parameters, to solve the issue of suboptimal model training performance caused by the lack of personalization in the traditional federated learning client model, which lays a theoretical foundation for the design and application of secure and efficient quantum federated learning algorithm. Furthermore, we design a personalized quantum federated learning algorithm, including aggregation, local training, and global updating, which ensures the security of data interaction between client and server and the privacy of client image data. In addition, we explore the application of the personalized quantum federated learning algorithm in the privacy image classification task on the FashionMNIST dataset.
	The main contributions of this work are summarized as follows.
	\begin{itemize}
		\item A personalized quantum federated learning model is proposed to achieve personalization in client model during the global training process without the need for additional local training.
		
		\item A personalized quantum federated learning algorithm is designed, using quantum channels to securely aggregate model parameters and protect the privacy of client image data during server-client interactions.
		
		\item The image classification experiments carried out on the FashionMNIST dataset demonstrate that our PQFL method shows superior performance compared with the non-personalized quantum federated learning method.
	\end{itemize}
	
	The outline of the paper is listed as follows. In Section \ref{section_ii}, we review the related works to quantum computing, quantum neural network and federated learning. Then, a personalized quantum federated learning method is proposed in Section \ref{section_iii}. Section \ref{section_iv} presents the experimental setup and analyzes the performance and security of the proposed method. And the conclusion of our work is provided in the last section.

\section{Related work} \label{section_ii}
	\subsection{Preliminary}	
		\subsubsection{Qubit}
			Different from the classical bit, a quantum bit (qubit) has the distinguishing characteristic of existing in a superposition state of $|0\rangle=[1,0]^T$ and $|1\rangle=[0,1]^T$ \cite{Nielsen_Chuang_2010}, which can be noted as follows.
			\begin{equation}
				|\phi\rangle = \alpha |0\rangle + \beta |1\rangle,
			\end{equation}
			where $|\alpha |^2 + |\beta |^2 = 1$. In addition, $|\alpha|^2$ and $|\beta|^2$ are the occurrence probability of state $|0\rangle$ and $|1\rangle$, respectively.				
			\subsubsection{Evolution}
			The quantum system always changes its state over time, which is called as evolution. The evolution process can be mathematically described as:
			\begin{equation}
				|\psi_2 \rangle = U |\psi_1\rangle,
			\end{equation}
			where $U$ is a unitary matrix satisfying $U^\dagger U=I$, also called quantum gates. Table \ref{table:QuantumOperator} presents the quantum gates that are commonly used.
			\begin{table}[h]
				\renewcommand{\arraystretch}{1.3}
				\begin{center}
					\caption{Commonly used quantum unitary operations.}
					\label{table:QuantumOperator}
					\begin{tabular}{ c  c  c }
						\hline
						{Unitary operation} & {Graph representation}& {Matrix representation}	\\
						\hline
						{H}  & {\begin{minipage}{0.3\columnwidth}\Qcircuit @C=1em @R=1em {&\gate{H} & \qw }\end{minipage}} & {$\frac{\sqrt{2}}{2}\begin{pmatrix} 1 & 1 \\ 1 & -1 \end{pmatrix}$} \\
						{X}  & {\begin{minipage}{0.3\columnwidth}\Qcircuit @C=1em @R=1em {&\gate{X} & \qw }\end{minipage}} & {$\begin{pmatrix} 0 & 1 \\1 & 0 \end{pmatrix}$} \\
						{Z}  & {\begin{minipage}{0.3\columnwidth}\Qcircuit @C=1em @R=1em {&\gate{Z} & \qw }\end{minipage}} & {$\begin{pmatrix} 1 & 0 \\0 & -1 \end{pmatrix}$} \\
						{RY}  & {\begin{minipage}{0.3\columnwidth}\Qcircuit @C=1em @R=1em {&\gate{RY(2\theta)} & \qw }\end{minipage}} & {$\begin{pmatrix} cos\theta & -sin\theta \\sin\theta & cos\theta \end{pmatrix}$} \\
						{CNOT}  & {\begin{minipage}{0.3\columnwidth}\Qcircuit @C=1em @R=1em {& \ctrl{0} \qwx[1] & \qw \\& \targ & \qw }\end{minipage}} & {$\begin{pmatrix} 1 & 0 & 0 & 0 \\0 & 1 & 0 & 0\\0 & 0 & 0 & 1\\0 & 0 & 1 & 0\\ \end{pmatrix}$} \\
						\hline
					\end{tabular}
				\end{center}
			\end{table}
		
		\subsubsection{Measurement}
		Measurement representing a non-unitary operation is an irreversible process, which is represented by a group of measure operators ${M_m}$, satisfying $\sum_m M^\dagger_m M_m = I$. Suppose the quantum system state is $|\psi \rangle$ before the measure operation, and the probability of measured result $m$ can be calculated as follows:
		\begin{equation}
			p(m) = \langle \psi | M^\dagger_m M_m | \psi \rangle.
		\end{equation}
		The sum of the measured probabilities $\sum_m p(m)$ equals 1.
		After measuring the system state, the quantum state collapses into:
		\begin{equation}
			|\psi ^\prime \rangle = \frac{M_m | \psi \rangle} {\sqrt{p(m)}}. 
		\end{equation}
		
		In quantum machine learning \cite{QSAN}, Pauli measurement is usually employed to determine the expected value of the current quantum state, and its measurement operator is the Pauli operator like $X$, $Y$, and $Z$ gate. The expected value can be calculated as:
		\begin{equation}
			E = \langle\psi|Hams|\psi\rangle.
		\end{equation}
		where $Hams$ is a Hamiltonian measured by Pauli on certain qubits.
		
	\subsection{Quantum Neural Network}
\begin{figure}[htbp]
			\centering
			\includegraphics[width=0.4\textwidth]{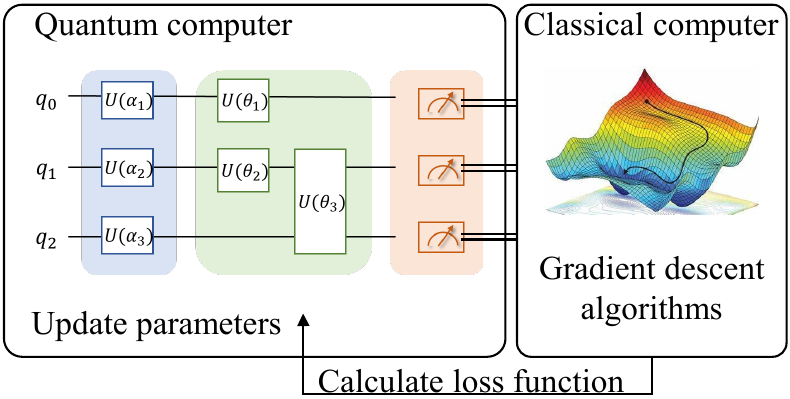}
			\caption{Quantum neural network with classical optimization.}
			\label{fig:qnn}  
		\end{figure}

		A Quantum Neural Network (QNN) \cite{qnc1995}, also called a parameterized quantum circuit built on a recent quantum computer, with its parameters optimized on the classical computer, which is shown in Fig. \ref{fig:qnn}. 
		The quantum circuit of QNN can be represented as a series of unitary gates:
		\begin{equation}
			U_{QNN}(\boldsymbol{\alpha},\boldsymbol{\theta}) = U_{ansatz}(\boldsymbol{\theta})U_{encoder}(\boldsymbol{\alpha}).
		\end{equation}
		
		In 2018, Farhi \textit{et al.} \cite{Farhi2018} proposed a universal realizable scheme for quantum neural networks to deal with classical or quantum classification problems. In 2019, Cong \textit{et al.} \cite{cong2019} introduced quantum convolutional neural networks, drawing inspiration from traditional convolutional neural networks. In 2020, Shi \textit{et al.} \cite{cvqnn2020} proposed an efficient cryptographic scheme based on continuously variable quantum neural networks, which is regarded as an efficient quantum cryptographic scheme and has potential application to quantum devices \cite{qcnn2022}. In 2021, a hybrid quantum-classical Convolutional Neural Network (CNN) was proposed by Liu \textit{et al.} \cite{liu2021}, which demonstrated superior learning accuracy in classification tasks compared to a classical CNN with an identical structure. Shi \textit{et al.} \cite{yongze} developed a quantum circuit learning algorithm utilizing parameterized Boson sampling to achieve fast fitting of Gaussian functions. In 2022, Wang \textit{et al.} \cite{wang2022} introduced the Quantum Pulsed Coupled Neural Network (QPCNN), building on the Piecewise Convolutional Neural Network (PCNN), and developed a quantum image segmentation algorithm using this model. Their analysis demonstrates that QPCNN offers exponential speedup compared to the classical PCNN.	
		In 2023, Shi \textit{et al.} \cite{WenXuan} introduced a parameterized Hamiltonian learning method utilizing quantum circuits, offering a novel avenue for addressing practical application challenges in quantum devices. These studies help to solve increasingly complex problems, providing support for quantum federated learning.
		
	\subsection{Federated Learning} 
		Federated learning enables collaborative training among users without exposing local data, resulting in a global machine-learning model. In recent years, research on FL has emerged in an endless stream. J. Konecny \textit{et al.} \cite{Jakub2016} initially introduced the idea of federated learning in 2016, and a federated optimization algorithm for distributed data was proposed. In the next year, McMahan \textit{et al.} \cite{McMahan2017} introduced the well-known FedAvg algorithm, which established the foundation for FL. In 2023, Sun \textit{et al.} \cite{10201382} proposed an efficient federated learning method using an adaptive client optimizer in image classification. 
		In 2024, Peng \textit{et al.} \cite{peng2024fedpft} proposed federated proxy fine-tuning to enhance foundation model adaptation in downstream tasks through FL. Since then, numerous studies have been conducted, exploring various aspects and extensions of federated learning.
		Among these, personalized federated learning model has attracted attention due to its capability to tailor models to individual client needs, while quantum federated learning is emerging as an innovative approach that leverages quantum computing principles to improve the security and performance of FL.
		\subsubsection{Personalized Federated Learning}
			It can be broadly categorized into two types of personalized federated learning: personalized for the global model and personalized for the learning model. 
			The former means that training a global model first, then conducting additional local training on each client's dataset to personalize it. In 2020, Dinh \textit{et al.} \cite{Dinh2020} proposed a FL algorithm utilizing Moreau envelopes as the regularized loss functions for clients, which helps utilize the global model in FL to create a personalized model tailored to each client’s data. 
			In 2021, Acar \textit{et al.} \cite{Acar2021} trained a meta-model globally, which is later customized locally to suit every device's specific needs.
			
			In addition, personalized for the learning model means that creating tailored models by adjusting the FL model aggregation process, accomplished through various learning paradigms in FL. 
			In 2022, Ma \textit{et al.} \cite{9880164} introduced a layer-wise FL model, which identifies the significance of each layer from various local clients, thereby optimizing the aggregation for personalized client models with the heterogeneous data.
			In 2023, Li \textit{et al.} \cite{10130784} presented a new transformer-based FL framework that personalizes self-attention for clients, sharing and aggregating other parameters of clients.
			Xu \textit{et al.} \cite{xu2023personalized} introduced a personalized FL approach, where explicit local-global feature alignment is achieved by utilizing global semantic knowledge to enhance representation learning.
			
			The above studies indicate that personalized federated learning can solve the problem of poor convergence of traditional federated learning algorithms on highly heterogeneous data, which is also a challenge faced by QFL.
			
		\subsubsection{Quantum Federated Learning}
			In 2021, Li \textit{et al.} \cite{ Li2021 } developed a QFL scheme by utilizing the principles of blind quantum computing, preserving the security of private data and offering valuable guidance in exploring quantum advantages in machine learning, particularly from a security standpoint. In 2022, Huang \textit{et al.} \cite{ Huang2022}, influenced by the classical federated learning algorithm, proposed the efficient communication learning of variational quantum algorithms from scattered data, known as quantum federated learning, which inspired new research in secure quantum machine learning. In 2023, Waleed \textit{et al.} \cite{ Yamany2023} introduced an optimized QFL framework aimed at safeguarding intelligent transportation systems and offering enhanced protection against various types of adversarial attacks.
			
			However, current quantum federated learning algorithms rarely consider the personalization of federated learning models, especially in personalized learning models. Therefore, we propose a personalized quantum federated learning algorithm to improve the personalization of client model with non-IID images in QFL.
			
\section{Personalized Quantum Federated Learning}\label{section_iii}
	\subsection{PQFL Model}
\begin{figure*}[htbp]
			\centering
			\includegraphics[width=0.8\textwidth]{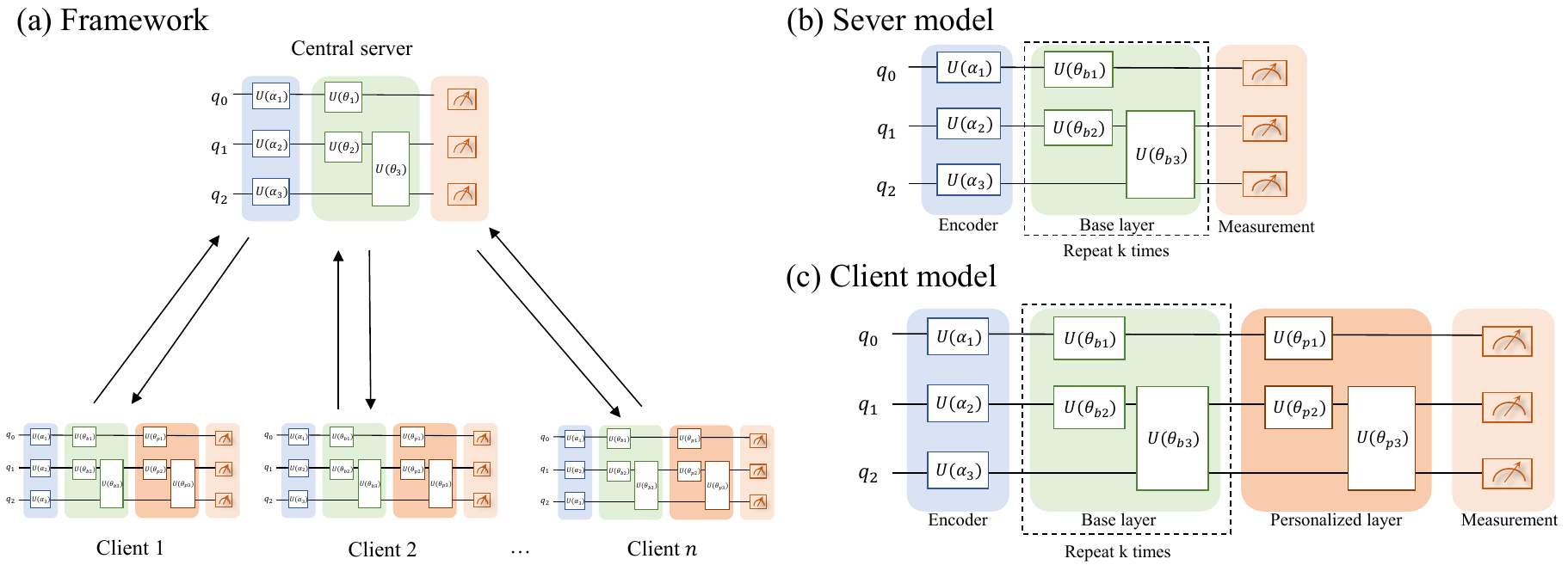}
			\caption{Personalized quantum federated learning framework.}
			\label{fig:pqfl}     
		\end{figure*}

		Fig. \ref{fig:pqfl} illustrates the basic architecture of the PQFL model, which includes a server and several clients, all of which have quantum computing capabilities.
		Firstly, the server model is a QNN containing an encoder $U_{encoder}(\vec{\alpha})$, an ansatz $U_{base}(\vec{\theta_b})$ where the base layer repeats $k$ times and measurement, which is shown in Fig. \ref{fig:pqfl} (b). The server model can be expressed as:
		\begin{equation}\label{Eq:pqfl_server_structure}
			\begin{aligned}
				U_{server}(\vec{\alpha}, \vec{\theta_b}) &= U_{base}(\vec{\theta_{b}})U_{encoder}(\vec{\alpha}).
			\end{aligned}
		\end{equation}
		The structure of the client model is consistent with that of the server, but an additional personalized layer $U_{personal}(\vec{\theta_p})$  as shown in Fig. \ref{fig:pqfl} (c) is added in the ansatz $U_{ansatz}(\vec{\theta})$. The client model is expressed as:	
		\begin{equation}\label{eq:qfl_client_structure}
			\begin{aligned}
				U_{client}(\vec{\alpha}, \vec{\theta}) &= U_{ansatz}(\vec{\theta})U_{encoder}(\vec{\alpha})\\&= U_{person}(\vec{\theta_{p}})U_{base}(\vec{\theta_{b}})U_{encoder}(\vec{\alpha}),
			\end{aligned}
		\end{equation}	
		where $\vec{\theta} = (\vec{\theta_b},\vec{\theta_p})$. All clients share the structure and parameters of the common base layer and have unique personalized layer parameters. 
		When training the model, the client updates the quantum circuit parameters for both the base and personalized layers simultaneously, but only the base layer parameters are uploaded to the server.
		
	\subsection{Quantum Circuit of PQFL}
\begin{figure}[htbp] 
			\centering
			\includegraphics[width=0.45\textwidth]{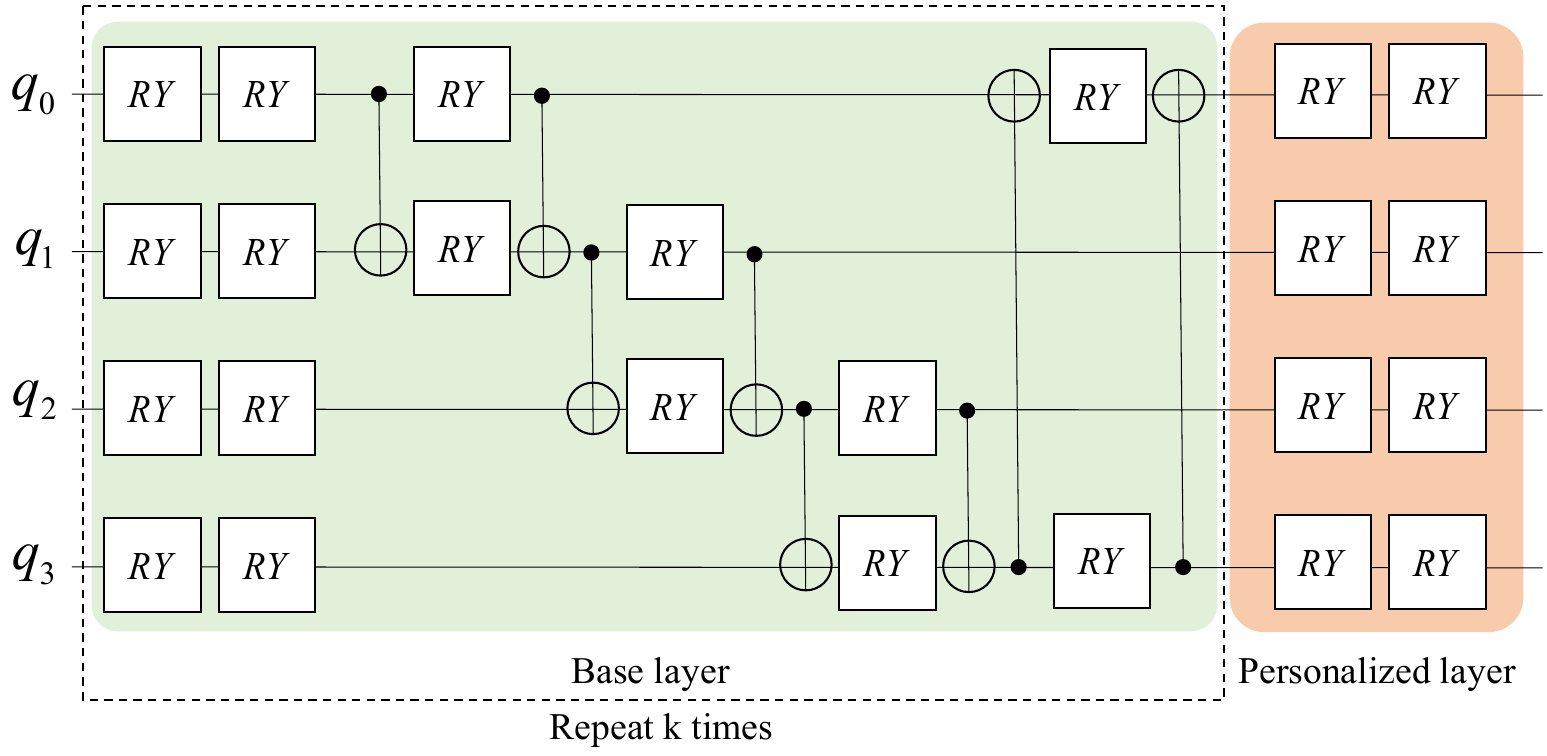}
			\caption{The ansatz quantum circuit of client.}
			\label{fig:pqfl_client_qc}  
		\end{figure}

		The client encodes the classical information into the amplitude of a quantum state. Fig. \ref{fig:encoder} illustrates the quantum circuit of the amplitude encoder $U_{encoder}(\vec{\alpha})$. Through amplitude encoding, an $N$-bit classical data is encoded into an $n$-qubit quantum state, where $n = \lceil \log_{2}N \rceil$. Usually, the amplitude encoder can be realized by using $X$, $RY$, and the controlled $RY$ gate. Fig. \ref{fig:pqfl_client_qc} shows the client model, whose quantum circuit of the base layer can be expressed as:
		\begin{figure}[htbp]
			\centering
			\includegraphics[width=0.45\textwidth]{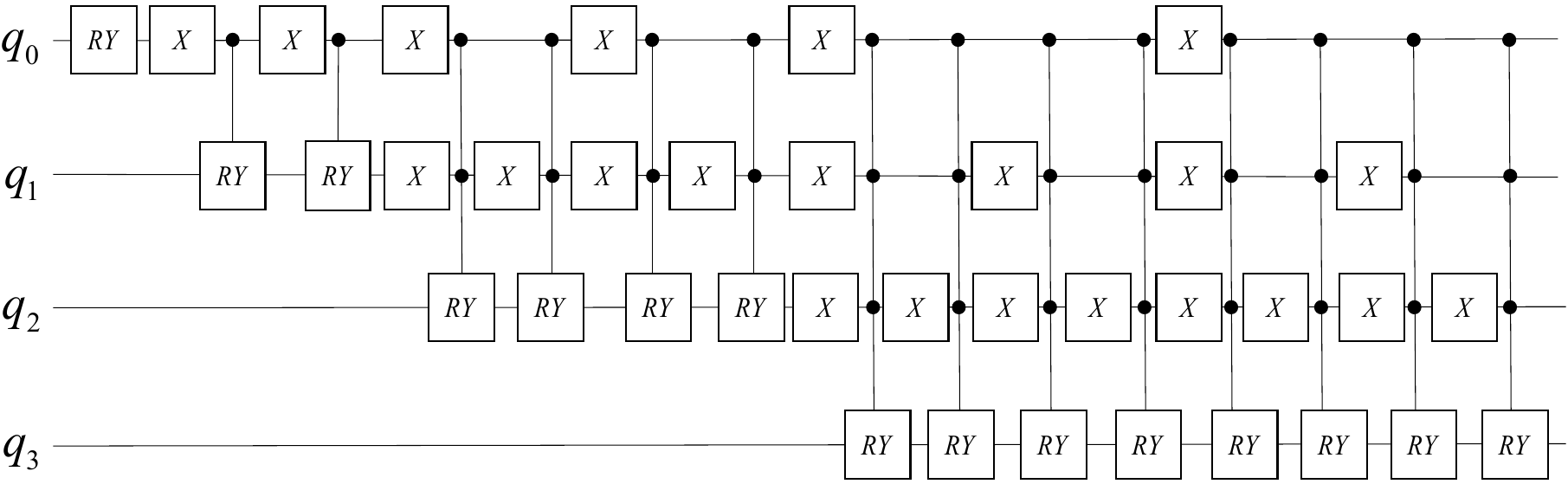}
			\caption{The quantum circuit of encoder of client.}
			\label{fig:encoder}  
		\end{figure}
		\begin{equation}\label{Eq:pqfl_base}
			\begin{aligned}
				U_{base}(\vec{\theta_b}) = &\otimes_{z=1}^k\{\otimes_{i=0,j=(i+1)\% n}^{n-1}[CNOT_{i,j}\\
				&(RY_j\otimes RY_{j+1}) CNOT_{i,j}]\\
				&\otimes_{i=0}^{n-1}(RY_{i}\otimes RY_{i})\},
			\end{aligned}
		\end{equation}
		where $CNOT_{i,j}$ indicates that the $j$th qubit is flipped if the $i$th qubit is 1; otherwise, no operation is taken. $RY_i$ denotes the rotation gate $RY$ acted on the $i$th qubit. The quantum circuit of the personalized layer can be expressed as:
		\begin{equation}\label{Eq:pqfl_person}
			\begin{aligned}
				U_{person}(\vec{\theta_p}) = \otimes_{i=0}^{n-1}(RY_{i}\otimes RY_{i}).
			\end{aligned}
		\end{equation}
		So the client model can be represented as:
		\begin{equation}\label{Eq:pqfl_client}
			\begin{aligned}
				U_{client}(\vec{\alpha}, \vec{\theta}) = &U_{person}(\vec{\theta_{p}})U_{base}(\vec{\theta_{b}})U_{encoder}(\vec{\alpha})\\
				= &\otimes_{i=0}^{n-1}(RY_{i}\otimes RY_{i})\prod \limits_{z=1}^k\{\otimes_{i=0,j=(i+1)\% n}^{n-1}\\
				&[CNOT_{i,j}(RY_j\otimes RY_{j+1}) CNOT_{i,j}]\\
				&\otimes_{i=0}^{n-1}(RY_{i}\otimes RY_{i})\}U_{encoder}(\vec{\alpha}),
			\end{aligned}
		\end{equation}
		The quantum circuit of the encoding and base layer of the server is consistent with that of the client, and the quantum circuit of the server can be expressed as:
		\begin{equation}\label{Eq:pqfl_server}
			\begin{aligned}
				U_{server}(\vec{\alpha}, \vec{\theta}) = &U_{base}(\vec{\theta_{b}})U_{encoder}(\vec{\alpha})\\
				= &\otimes_{z=1}^k\{\otimes_{i=0,j=(i+1)\% n}^{n-1}[CNOT_{i,j}\\
				&(RY_j\otimes RY_{j+1})CNOT_{i,j}]\\
				&\otimes_{i=0}^{n-1}(RY_{i}\otimes RY_{i})\},
			\end{aligned}
		\end{equation}
		
	\subsection{PQFL Algorithm}
		Under the assumption that the samples owned by the client are non-IID, and all participants are honest but curious, we introduce the personalized quantum federated learning algorithm. Before that, we will introduce the parameter aggregation algorithm, the local training algorithm.
		
		\subsubsection{Aggregation}
			Based on the quantum security aggregation protocol \cite{Zhang2022}, we improve a quantum parameter weighted average algorithm shown in Algorithm \ref{alg:qsa} for non-IID data, which can prevent disclosing any individual client parameter information.
			\begin{algorithm}[htbp]
				\small
				\caption{Quantum parameter weighted average algorithm.}
				\label{alg:qsa}
				\begin{algorithmic}[1]
					\STATE The client transmits the number of samples to the server through classical channel, and the server calculates the weighted score according to the Eq. (\ref{Eq:client_weight_frac}) and returns it to the client.
					\STATE The server generates $N$ GHZ states $|\varPsi \rangle = |\varPsi_0\rangle|\varPsi_1\rangle\cdots|\varPsi_N\rangle$.				
					\FOR{$i$ from $1$ to $N$}
					\STATE The server sends $M$ qubits of $|\varPsi_i \rangle$ to $M$ clients via quantum channel, respectively.
					\FOR{$m$ from $1$ to $M$}
					\STATE The client $C_m$ performs $RY(F_m\cdot\theta_{m,i})$ operation on the received qubit.
					\STATE The client sends the qubit to the server using quantum channel.
					\ENDFOR
					\STATE The server performs the decoding operation according to the Eq. (\ref{Eq:server_decode}).
					\STATE The server obtains the estimation of the $i$th parameter $\theta_{i} = \sum_{m= 1}^{m} (F_m\cdot\theta_{m,i}) $ by measurement.
					\ENDFOR\\
					The server calculates the weighted average parameters according to the Eq. (\ref{Eq:server_aggregate}) to obtain the aggregated parameters $\vec{\theta}$.
				\end{algorithmic}
			\end{algorithm} 
			
			Firstly, in each round of the global training, the server first chooses $M$ clients $C = \{C_1,C_2,\cdots,C_M\}$ to be trained. For the client $C_m$, it sends the training image number $l_m$ to the server through classical channel. After receiving $l_m$ of all the participant clients, the server calculates the weighted score of each client according to Eq. (\ref{Eq:client_weight_frac}), and then sends it to the client through the classical channel, so that each client can get its weighted score.
			\begin{equation}\label{Eq:client_weight_frac}
				F_m=\frac{l_m}{\sum_{i=1}^{M}l_i}.  
			\end{equation}
			Secondly, the server generates $N$ GHZ states \cite{GHZ} $|\varPsi \rangle$ as shown in Eq. (\ref{Eq:server_ghz}):
			\begin{equation}\label{Eq:server_ghz}
				\begin{aligned}
					|\varPsi \rangle &= |\varPsi_0\rangle|\varPsi_1\rangle\cdots|\varPsi_{N-1}\rangle,
				\end{aligned}
			\end{equation}
			where $|\varPsi_i\rangle=\frac{1}{\sqrt{2}}(|0\rangle^{\otimes M}+|1\rangle^{\otimes M})$, and $N$ is related to the number of the base layer parameters of the client QNN model. Then the server sends $M$ qubits of each GHZ state, which consists of $M$ qubits, to the $M$ participant clients through the quantum channel.
			When the client $C_m$ receives one of the qubits of the $i$th GHZ state $|\varPsi_i\rangle$, it encodes the qubit by applying a revolving gate $RZ(F_m\cdot\theta_{m,i})$ gate. Where $\theta_{m,i}$ represents the $i$th parameter in the base layer of the client $C_m$, and $F_m$ represents the weighted fraction of the client $C_m$. Other clients perform the same operation on the received qubits and send the encoded qubits back to the server. The $i$th base layer parameter of all clients is encoded into the GHZ state:
			\begin{equation}\label{Eq:client_encoding}
				|\varPsi'_i \rangle = \frac{1}{\sqrt{2}}(|00\cdots0\rangle+e^{i\sum_{m=1}^{M}(F_m\cdot\theta_{m,i})}|11\cdots1\rangle).
			\end{equation}
			The client $C_m$ then sends the encoded GHZ state $|\varPsi'_m\rangle$ to the server through the quantum channel. Then the server decodes the GHZ state, using $CNOT$ gate and $H$ gate to decode the entangled state $|\varPsi'_i\rangle$, then the quantum state evolves into:
			\begin{equation}\label{Eq:server_decode}
				\begin{aligned}  
					|\varPsi''_i\rangle&=H_1CNOT_{1,2}\cdots CNOT_{N-1,N}|\varPsi'_i\rangle \\
					&= \frac{1}{\sqrt{2}}[(1+e^{i\sum_{m=1}^{M}(F_m\cdot\theta_{m,i})})\\
					&|00\cdots0\rangle+(1-e^{i\sum_{m=1}^{M}(F_m\cdot\theta_{m,i})})|11\cdots1\rangle],
				\end{aligned}
			\end{equation}
			Then, as the server measures one qubit, the GHZ state will obtain $|0\rangle$ with the probability of $\text{Pr}=\frac{1}{2}(1+\cos(\sum_{m=1}^{M}(F_m\cdot\theta_{m,i}))$. By repeating this process, the server will get the estimations of $\sum_{m=1}^{M}(F_m\cdot\theta_{m,i})$, which is calculated as follows:
			\begin{equation}\label{Eq:server_estimate}
				\begin{aligned}
					\sum_{m=1}^{M}(F_m\cdot\theta_{m,i}) = \arccos(2\text{Pr}-1).
				\end{aligned}
			\end{equation}
			In this way, the server gets the weighted average of the $i$th parameter in the base layer of all clients:
			\begin{equation}\label{Eq:server_i_average}
				\begin{aligned}
					\theta_{i} = \sum_{m=1}^{M}(F_m\cdot\theta_{m,i}). 
				\end{aligned}
			\end{equation}
			In this training round, the client performs the same operations as above for other parameters in the base layer. Thus, the server gets the weighted average aggregated parameters of all clients:
			\begin{equation}\label{Eq:server_aggregate}
				\begin{aligned}
					\vec{\theta}&=(\theta_0,\theta_1,\cdots,\theta_N)\\
					&=(\sum_{m=1}^{M}(F_m\cdot\theta_{m,0}),\sum_{m=1}^{M}(F_m\cdot\theta_{m,1}),\cdots,\sum_{m=1}^{M}(F_m\cdot\theta_{m,N})),
				\end{aligned}
			\end{equation}
			where $\vec{\theta_{m}}$ represents the base layer parameters of client $C_m$.
		
		\subsubsection{Local Training}
			The local training algorithm can be described as Algorithm \ref{alg:client_update}.
			\begin{algorithm}[htbp]
				\small
				\caption{Local Training Algorithm}
				\label{alg:client_update}
				\begin{algorithmic}[1]
					\STATE Generate label category distribution $Q \leftarrow Dir(\alpha)$.
					\STATE Divide data samples to $M$ clients according to label category distribution $Q$.
					\STATE The client to initialize $U_ {base} (\vec {\theta_b}) $ and $U_ {person} (\vec {\theta_p})$.		
					\FOR{$e$ from $1$ to $E$}
					\STATE Encoding classical data to quantum states $|\psi\rangle$.
					\STATE Carry out $U_{person}(\vec{\theta_p})U_{base}(\vec{\theta_b})|\psi\rangle$.
					\STATE Measuring the output quantum state $|\varphi\rangle$ with the operator $H_{hams}$, and calculating the expectation value according to Eq. (\ref{Eq:expect_client}).
					\STATE Calculating the loss function $\ell_m (\theta_ {t}) $ according to Eq. (\ref{Eq:loss_client}).
					\STATE Optimizing parameters $\vec{\theta_b}$ and $\vec{\theta_p}$ with Adam algorithm.
					\ENDFOR
				\end{algorithmic}
			\end{algorithm} 			
			Dirichlet distribution is used to simulate the non-IID samples of the clients. The probability density function of the Dirichlet distribution is defined as follows:
			\begin{equation}\label{Eq:dirichlet}
				f(\theta_1,...,\theta_L;\alpha_1,...,\alpha_L)=\frac{1}{B(\boldsymbol{\alpha})}\prod_{i=1}^L\theta_i^{\alpha_i-1},
			\end{equation}
			where $\theta_i\geq0$ and $\sum_{i=1}^{L}\theta_i=1$. $B(\boldsymbol{\alpha})$, which is a multivariate Beta function, is defined as follows:
			\begin{equation}\label{Eq:beta}
				B(\boldsymbol{\alpha})=\frac{\prod_{i=1}^L\Gamma(\alpha_i)}{\Gamma(\sum_{i=1}^{L}\alpha_i)}, \alpha=(\alpha_1,...,\alpha_L).
			\end{equation}
			Particularly, when $\alpha_i$ in vector $\boldsymbol{\alpha}$ has the same value, the distribution is a symmetric Dirichlet distribution, whose density function can be expressed as:
			\begin{equation}\label{Eq:symetry_dirichlet}
				f(\theta_1,...,\theta_L;\alpha)=\frac{\Gamma(\alpha L)}{\Gamma(\alpha)^L}\prod_{i=1}^L\theta_i^{\alpha-1}.
			\end{equation}
			A distribution $D$ of label categories of data samples satisfy the Dirichlet distribution $D\sim Dir(\alpha)$, which can be expressed as:
			\begin{equation}\label{Eq:dirichlet_label}
				D =
				\begin{pmatrix}
					d_{0,0}     & d_{0,1}     & \dots   &d_{0,M-1}\\
					d_{1,0}     & d_{1,1}     & \dots   &d_{2,M-1}\\
					\vdots      & \vdots      & \ddots  &\vdots\\
					d_{Y-1,0}   & d_{Y-1,1}   & \dots   &d_{Y-1,M-1}\\
				\end{pmatrix}_{Y\times M},
			\end{equation}
			where $d_{i,j}$ represents the distribution probability of the data samples of category $i$ on the $j$th client. According to the label category distribution matrix $D$, the data samples of the corresponding category are divided into $M$ clients. The distribution parameter $\alpha$ controls the similarity of sample distribution. As the distribution parameter $\alpha$ decreases, the disparity in the category distribution of client data samples increases. In extreme cases, each client contains data samples with only one label category. The larger the value of the distribution parameter $\alpha$, the more similar the category distribution of client data samples, and the closer it is to uniform distribution. For example, for $M$ clients and a dataset of $Y$ categories, the label category distribution matrix of $Y\times M$ is obtained from the Dirichlet distribution, and the samples of each category label are divided into corresponding clients according to different proportions in Eq. (\ref{Eq:dirichlet_label}).

			Then, the client trains on the divided data samples. All participants initialize the base layer $U_{base}(\vec{\theta_b})$, meanwhile the parameters in personalized layer $U_{person}(\vec{\theta_p})$ are initialized locally. 
			The loss function of client $C_m$ can be defined as:
			\begin{equation}\label{Eq:loss_client}
				\ell_m(\vec{\theta}) = - \frac{1}{N}\sum_{i=0}^{N-1}\log(\frac{\exp(E_i[c](\vec{\theta}))}{\sum_j^b \exp(E_i[j](\vec{\theta}))}),
			\end{equation}
			where $N$ is the batch size, $c$ is the index of ``1" in the one-hot encoding of the label, and $E_i$ is the expected value measured by the client, which is calculated as follows:
			\begin{equation}\label{Eq:expect_client}
				\begin{aligned}
					E_i(\vec{\theta})&=\langle\varphi|U_{ansatz}^\dagger(\vec{\theta})H_{hams}U_{ansatz}(\vec{\theta})|\psi\rangle\\
					&=\langle\varphi|U_{base}^\dagger(\vec{\theta_b})U_{person}^\dagger(\vec{\theta_p})H_{hams}\\
					&U_{person}(\vec{\theta_p})U_{base}(\vec{\theta_b})|\psi\rangle.
				\end{aligned}
			\end{equation}
			Adam algorithm is adopted to optimize both the parameters $\vec{\theta_b}$ and $\vec{\theta_p}$. After several rounds of training, the client transmits all of the base layer parameters to the server, while retaining the personalized layer parameters.
			
			\begin{figure*}[htpb]
				\centering
				\includegraphics[width=0.6\textwidth]{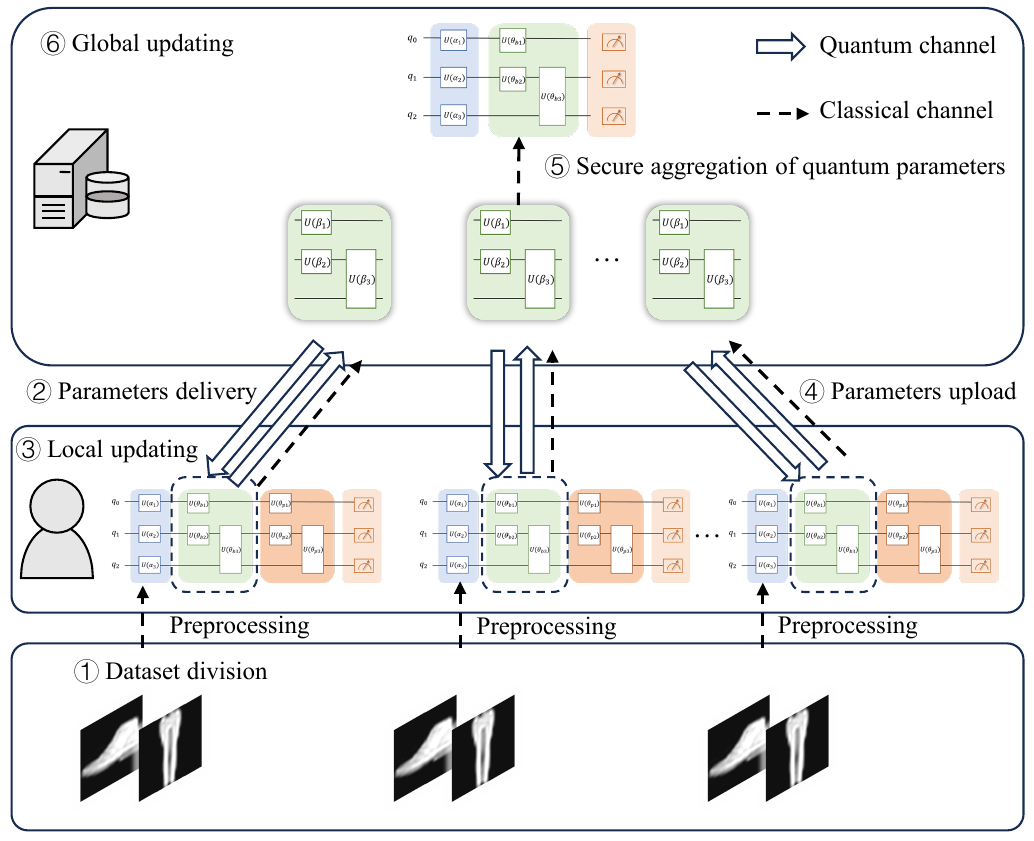}
				\caption{Personalized quantum federated learning for image classification.}
				\label{fig:pqfl_process}  
			\end{figure*}
			\subsubsection{Global Updating}
				Based on the aggregation algorithm and local training algorithm, we propose the PQFL algorithm in the global updating process, as shown in Algorithm \ref{alg:server_update}.
				The global objective function is:
				\begin{equation}\label{Eq:global_obj}
					\begin{aligned}
						f(\boldsymbol{\theta^*})=\mathop{min}\limits_{\boldsymbol{\theta}}f(\boldsymbol{\theta})=\mathop{min}\limits_{\boldsymbol{\theta}}\frac{1}{M}\sum_{m=1}^{M}\ell_m(\boldsymbol{\theta}),
					\end{aligned}
				\end{equation}
				where $M$ is the number of participant clients.
				
				\begin{algorithm}[htbp]
					\small
					\caption{PQFL algorithm}
					\label{alg:server_update}
					\begin{algorithmic}[1]
						\STATE The server initializes the base layer parameters $\vec{\theta_{b}}$,  simultaneously the client initializes the personalized layer parameter $\vec{\theta_{ps}}$.
						\FOR{$r$ from $1$ to $R$} 
						\STATE The server selects $M$ clients;
						\STATE The server sends the parameter $\vec{\theta_{b}}$ to $M$ clients through the quantum channels.
						\FOR{$e$ from $1$ to $E$}
						\STATE The clients receive the parameter $\vec{\theta_{b}}$, and stitch $\vec{\theta_{p}}$ into a vector.
						\STATE The clients optimize the parameters $\vec{\theta}=(\vec{\theta_b},\vec{\theta_p})$ according to Algorithm \ref{alg:client_update}.					
						\ENDFOR
						\STATE The clients send $\vec{\theta_b}$ to the server by quantum channels.
						\STATE The server aggregates parameters $\vec{\theta_{b}}$ according to Algorithm \ref{alg:qsa}.
						\ENDFOR
					\end{algorithmic}
				\end{algorithm}
				Firstly, the server initializes a set of base layer parameters $\vec{\theta_{b}}$, selects a set of clients $C=\{C_1,C_2,\cdots,C_M\}$, and the initialized parameters are sent to the clients through quantum channels. The server generates $M$ qubits:
				\begin{equation}\label{Eq:down_server}
					|\varphi\rangle=|\varphi_0\rangle|\varphi_1\rangle\cdots|\varphi_{M-1}\rangle,
				\end{equation}
				where $|\varphi_i\rangle=\frac{1}{\sqrt{2}}(|0\rangle+|1\rangle)$. The server encodes the $i$th parameter $\theta_{i}$ to $N$ qubits, that is, applies the rotation gate $RZ(\theta_{i})$ gate to the $N$ qubits, which is given by:
				\begin{equation}\label{Eq:down_server_encode}
					\begin{aligned}
						|\varphi'\rangle&=|\varphi_0'\rangle|\varphi_1'\rangle\cdots|\varphi'_{M-1}\rangle\\
						&=RZ(\theta_{i})|\varphi_0\rangle RZ(\theta_{i})|\varphi_1\rangle\cdots RZ(\theta_{i})|\varphi_{M-1}\rangle,
					\end{aligned}
				\end{equation}
				where $|\varphi_i'\rangle=\frac{1}{\sqrt{2}}(|0\rangle+e^{i\theta_i}|1\rangle)$. The server then sends the $M$-qubit $|\varphi'\rangle$ to $M$ clients through quantum channels, and the clients measure the received qubits and the quantum state will collapse to 0 with the probability of $\text{Pr}=\frac{1}{2}(1+\cos\theta_i)$. By repeating this process, the client will get an estimation of $\theta_i$, which is calculated as follows:
				\begin{equation}\label{Eq:client_estimate}
					\begin{aligned}
						\theta_i = \arccos(2\text{Pr}-1).
					\end{aligned}
				\end{equation}
				Other parameters of the base layer can also be sent to the client through the same process so that all clients can receive the parameter $\vec{\theta_b}$ from the server.
				
				Secondly, the clients update both base and personalized layer parameters according to Algorithm \ref{alg:client_update}, and upload the updated base layer parameter $\vec{\theta_{b}^t}$ to the server through Algorithm \ref{alg:qsa}. And the server calculates the weighted average sum of the parameters uploaded by each client according to  Eq. (\ref{Eq:server_aggregate}), and obtains a new model parameter $\vec{\theta_b^{t+1}}$.
				
				By repeating the above steps, the server can complete global updating.
				
				\subsection{PQFL for Image Classification}
					A privacy image classification scheme based on personalized quantum federation learning is proposed, including dataset partitioning, parameters delivery, local updating, parameters upload, secure aggregation of quantum parameters, and global updating, which is shown in Fig. \ref{fig:pqfl_process}. The server and clients are capable of quantum computing. The specific steps are as follows.
					
					\subsubsection{Dataset Partitioning}
\begin{figure*}[htbp]
						\centering
						\includegraphics[width=0.3\textwidth]{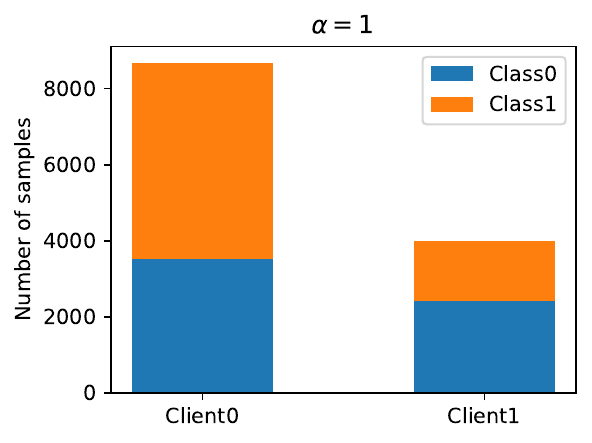}
						\includegraphics[width=0.3\textwidth]{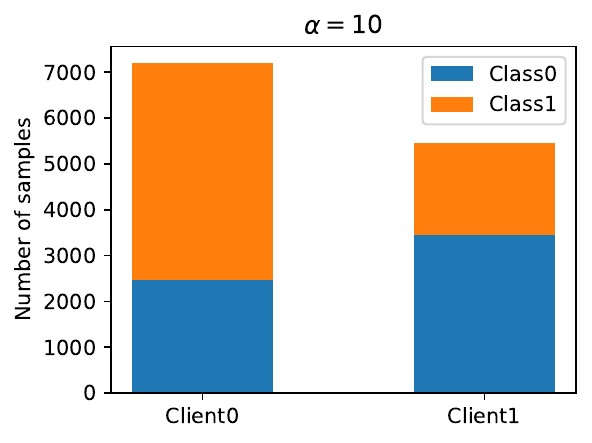}
						\includegraphics[width=0.3\textwidth]{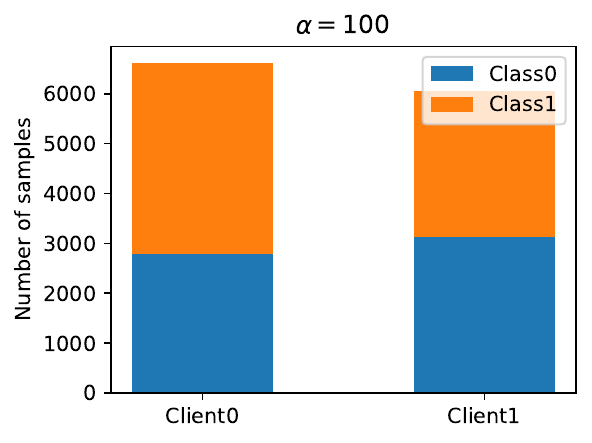}\\
						(a)\qquad\qquad\qquad\qquad\qquad\qquad\qquad(b)\qquad\qquad\qquad\qquad\qquad\qquad\qquad(c)\\
						\includegraphics[width=0.3\textwidth]{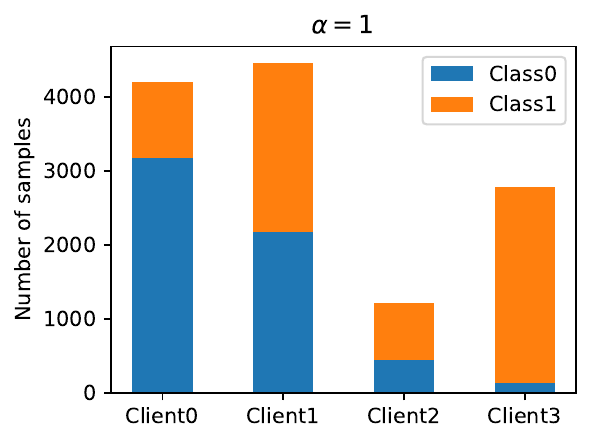}
						\includegraphics[width=0.3\textwidth]{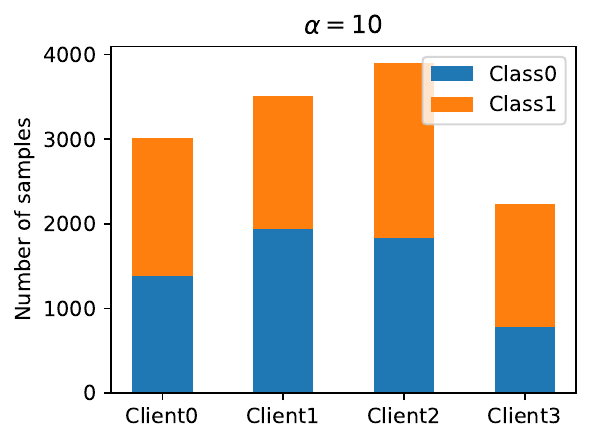}
						\includegraphics[width=0.3\textwidth]{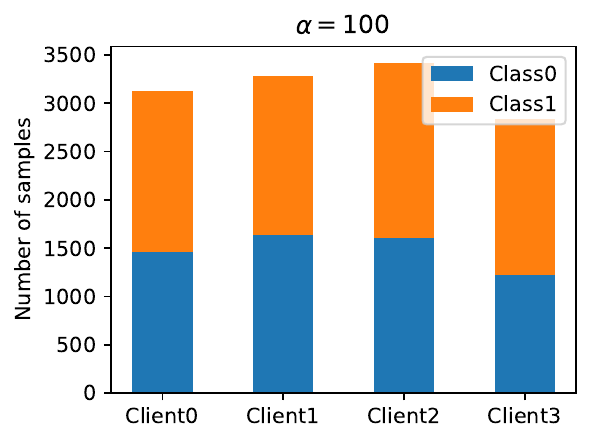}\\
						(d)\qquad\qquad\qquad\qquad\qquad\qquad\qquad(e)\qquad\qquad\qquad\qquad\qquad\qquad\qquad(f)\\
						\includegraphics[width=0.3\textwidth]{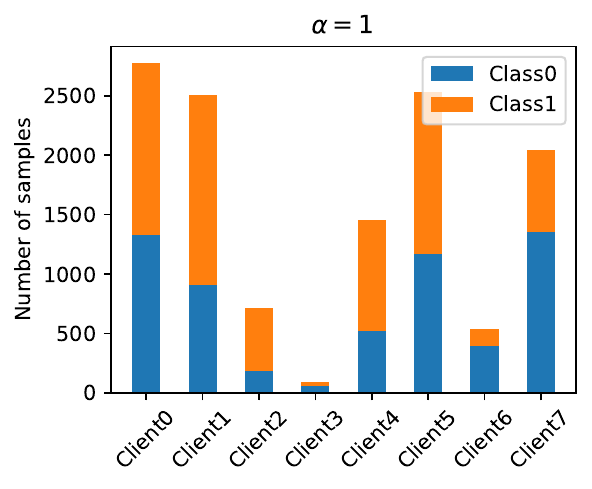}
						\includegraphics[width=0.3\textwidth]{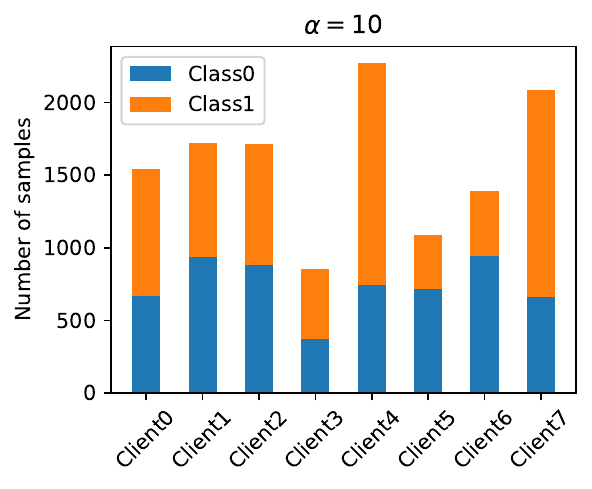}
						\includegraphics[width=0.3\textwidth]{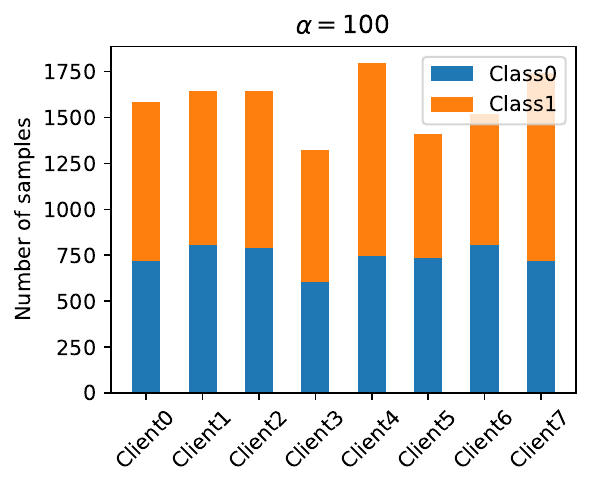}\\
						(g)\qquad\qquad\qquad\qquad\qquad\qquad\qquad(h)\qquad\qquad\qquad\qquad\qquad\qquad\qquad(i)
						\caption{Partitioning of non-IID images when $\alpha=1,10,100$. (a)$\sim$(c) is the data distribution of the two clients.  (d)$\sim$(f) is the data distribution of the four clients. (g)$\sim$(i) is the data distribution of 8 clients.}
						\label{fig:noniid_mnist}
					\end{figure*}

					The dataset is divided into non-IID according to the Dirichlet distribution and distributed to each client. For two classes of images, when the distribution parameter is $\alpha=1,10,100$, and the number of clients is 2, 4, and 8, the partition of non-IID images is shown in Fig. \ref{fig:noniid_mnist}.
					
					\subsubsection{Parameters Delivery}			
					The server constructs a 4-qubit quantum neural network model as shown in Fig. \ref{fig:pqfl} (b). The server initializes the base layer parameters $\vec{\theta_b}$, encodes the parameters into the quantum state according to Algorithm \ref{alg:server_update}, and sends the qubits to each client through the quantum channels. After receiving qubits containing parameter information, the clients measure the probability to get an estimation of the parameter and use it as a parameter in the base layer. In addition, the server calculates the weight score of each client through the classical channels and sends it to the corresponding client.
					
					\subsubsection{Local Training} 			
					The client builds a quantum neural network model containing 4 qubits, as shown in Fig. \ref{fig:pqfl} (c) and Fig. \ref{fig:pqfl_client_qc}. The base layer repeats $k=3$ times. Both the base layer and the personalized layer parameters are trained. The clients conduct several rounds of training updates to the local model according to Algorithm \ref{alg:client_update}, and obtain the trained parameter $\vec{\theta}$.
					
					\subsubsection{Parameters Upload}
					The clients send the number of image samples to the server through the classical channels. The server calculates the weight score of each client according to the number of image samples provided by each client.
					
					\subsubsection{Secure Aggregation of Quantum Parameters}
					The clients receive the weight score, multiply it with the base layer parameters $\vec{\theta_b}$, encodes it into the quantum state according to Algorithm \ref{alg:qsa}, and transmit it to the server through quantum channels. The weighted average of the base layer parameters $\vec{\theta_b}$ is calculated based on the server's measurements.
					
					\subsubsection{Global updating} 		
					The server updates the model parameters to weighted average parameters $\vec{\theta_b}$ of the base layer, repeating the above steps until reaching the number of training rounds or classification accuracy.
					
\section{Experiments and Analysis}\label{section_iv}

	\subsection{Experiment setting}
\begin{figure}[htbp]
			\centering
			\includegraphics[width=0.45\textwidth]{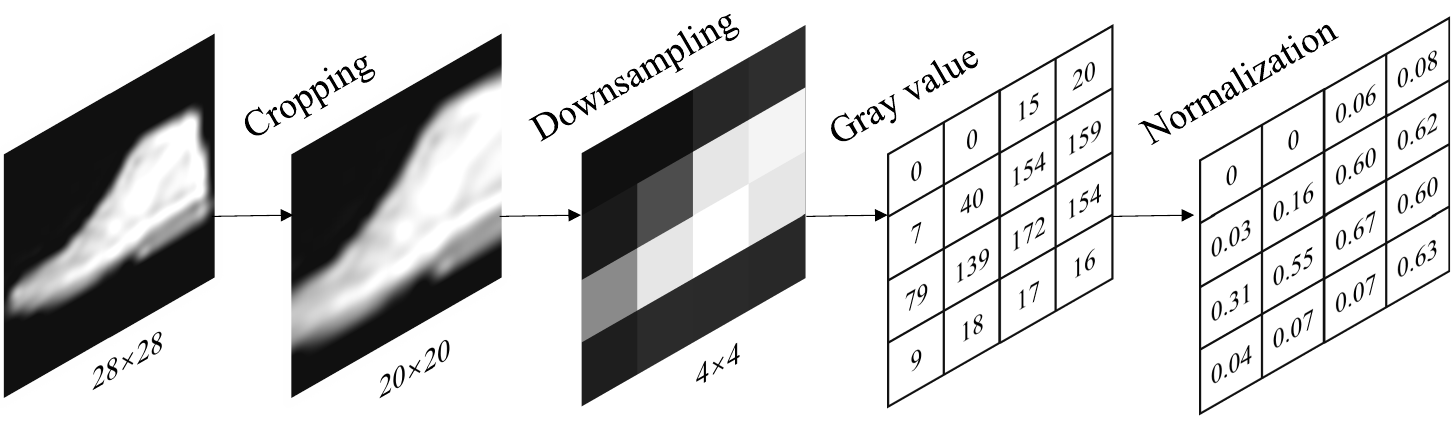}
			\caption{Preprocessing of image data.}
			\label{fig:preprocess}
		\end{figure}

		The experiments are conducted using the hardware device of Intel Core i7-9700 and the Mindspore quantum library is used for the experiment. The experimental outcomes indicate that the personalized quantum federated learning algorithm we proposed is capable of achieving high accuracy in binary classification tasks involving clothing images while securing the privacy images and models. Next, the accuracy, security, and communication overhead of private image classification schemes based on personalized quantum federated learning are analyzed. The hyperparameter settings of the experiment are shown in Table \ref{table:pqfl_hyperparameter}. For concreteness, we conduct a classification task about classifying images ``trouser or ankle boot" in the FashionMNIST dataset, which is a clothing dataset consisting of 10 categories with a total of 70000 grayscale images and 60,000 samples for training and 10,000 samples for testing. An original $28\times28$ image is preprocessed into a $4\times4$ image according to Fig. \ref{fig:preprocess}.
		\begin{table}[htbp]
			\begin{center}  
				\caption{Hyperparameter settings}
				\label{table:pqfl_hyperparameter}
				\setlength{\tabcolsep}{5mm}
				\renewcommand{\arraystretch}{1}
				\begin{tabular}{ c  c  c  c }
					\toprule
					Number of clients&2&4&8\\	
					\midrule
					Number of base layer&3&3&3\\
					Learning rate&0.01&0.01&0.01\\
					Batch size&50&50&50\\
					Global training rounds&100&100&100\\
					Local training rounds&1&1&1\\
					Optimizer&Adam&Adam&Adam\\
					\bottomrule
				\end{tabular}
			\end{center}
		\end{table}
		
		\subsection{Accuracy Analysis}
		We perform the training process using Dirichlet distribution parameters $\alpha = 1, 10, 100$. 
		
		Table \ref{table:pqfl_acc_server} shows that the server model in PQFL can achieve higher accuracy on the FashionMNIST dataset than the model without personalized layer in most settings. In particular, the model accuracy of the server can achieve 100\% when the distribution parameter $\alpha=100$ and the clients quantity is 8. As the distribution parameters and the clients quantity increases, the server model's accuracy remains relatively stable, indicating that the proposed PQFL model is robust to variations in distribution parameters and clients quantity.
		\begin{table}[htbp]
			\begin{center}  
				\caption{The model accuracy of the server.}
				\label{table:pqfl_acc_server}
				\setlength{\tabcolsep}{2mm}
				\renewcommand{\arraystretch}{1}
				\begin{threeparttable}
					\begin{tabular}{ c   c  cc}
						\toprule 
						$\alpha$ & $M$ & {With personalized layer} & {Without personalized layer}\\
						\midrule
						\multirow{3}{*}{1} &2&0.92&0.94\\
						&4&\textbf{0.99}&0.97	\\
						&8&\textbf{0.99}&0.95	\\
						\midrule
						\multirow{3}{*}{10} &2&0.97&0.98\\
						&4&0.94&0.97	\\
						&8&\textbf{0.97}	&0.95\\
						\midrule
						\multirow{3}{*}{100} &2&\textbf{0.98}&0.98\\
						&4&\textbf{0.97}&0.97	\\
						&8&\textbf{1.00}&0.93	\\
						\bottomrule
					\end{tabular}
					\begin{tablenotes}
						\footnotesize
						\item $M$ is the number of clients.
					\end{tablenotes}
				\end{threeparttable}
			\end{center}
		\end{table}
		
\begin{figure*}[htbp]
			\centering
			\includegraphics[width=0.3\textwidth]{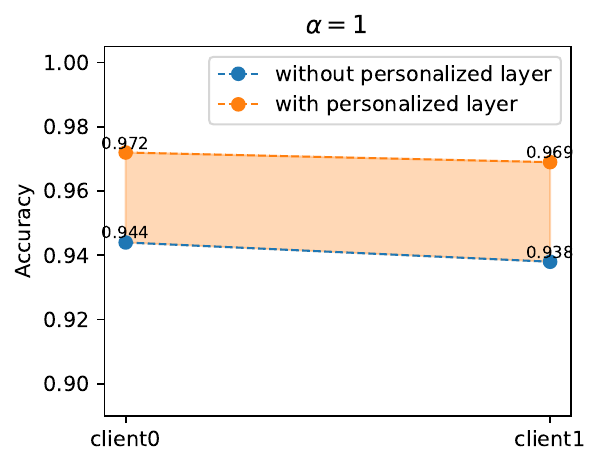}
			\includegraphics[width=0.3\textwidth]{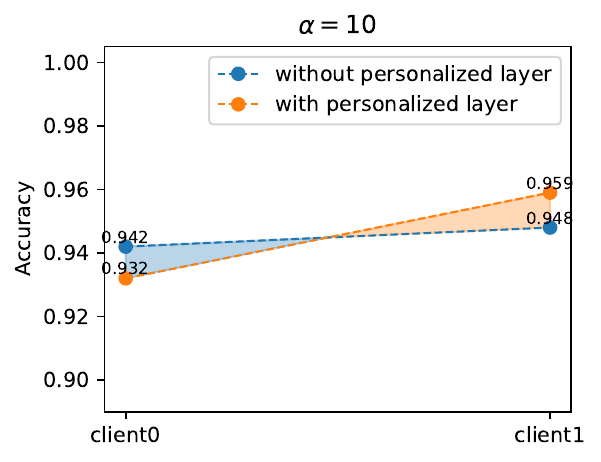}
			\includegraphics[width=0.3\textwidth]{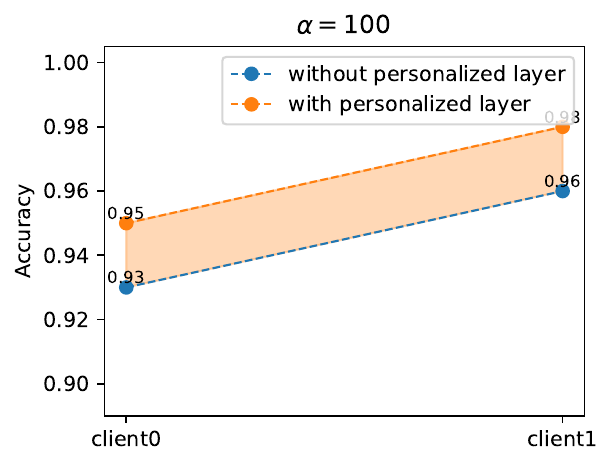}\\
			(a)\qquad\qquad\qquad\qquad\qquad\qquad\qquad(b)\qquad\qquad\qquad\qquad\qquad\qquad\qquad(c)\\
			\includegraphics[width=0.3\textwidth]{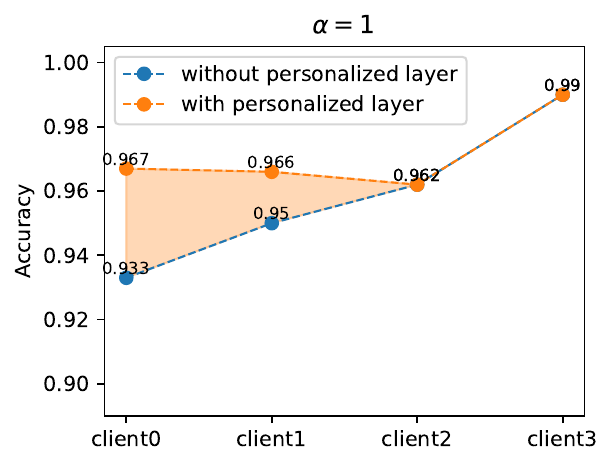}
			\includegraphics[width=0.3\textwidth]{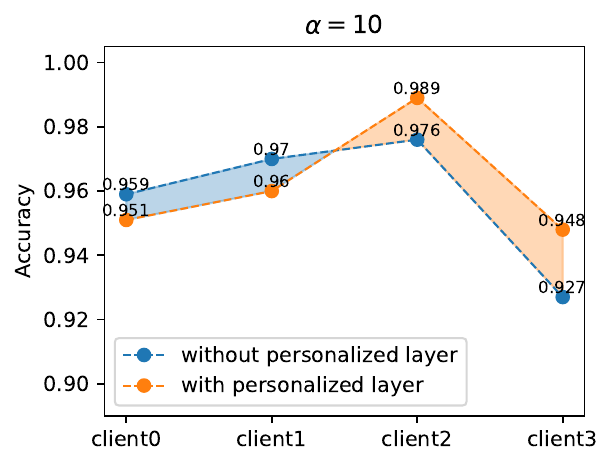}
			\includegraphics[width=0.3\textwidth]{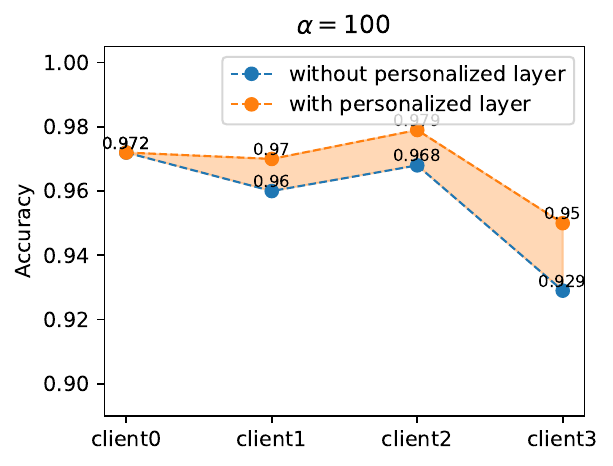}\\
			(d)\qquad\qquad\qquad\qquad\qquad\qquad\qquad(e)\qquad\qquad\qquad\qquad\qquad\qquad\qquad(f)\\
			\includegraphics[width=0.3\textwidth]{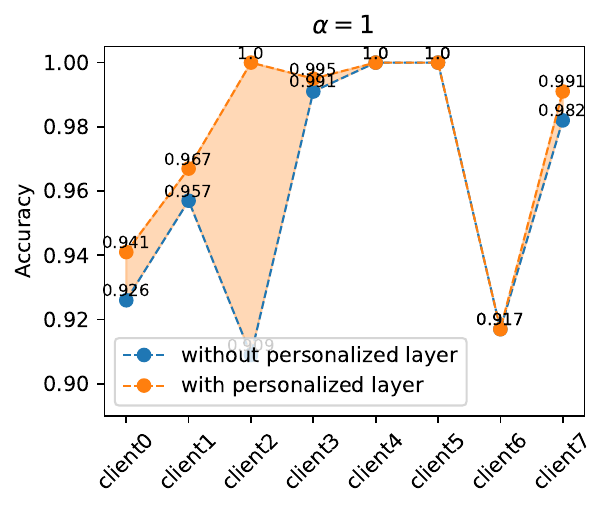}
			\includegraphics[width=0.3\textwidth]{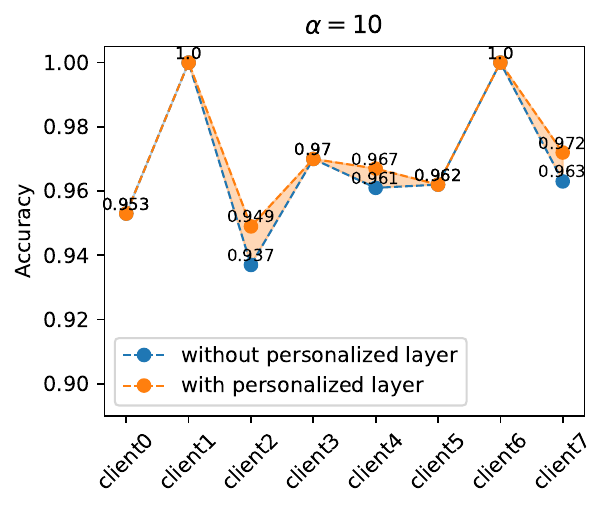}
			\includegraphics[width=0.3\textwidth]{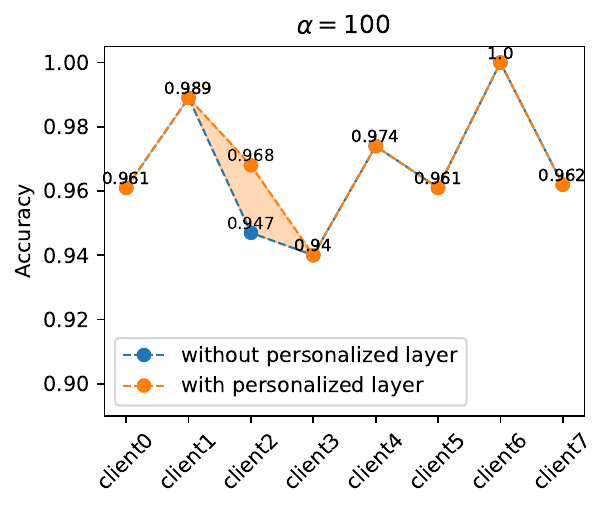}\\
			(g)\qquad\qquad\qquad\qquad\qquad\qquad\qquad(h)\qquad\qquad\qquad\qquad\qquad\qquad\qquad(i)
			\caption{Accuracy of client model with and without personalized layer on FashionMNIST dataset when $\alpha=1,10,100$, respectively. (a)$\sim$(c) is the accuracy of 2 clients. (d)$\sim$(f) is the accuracy of 4 clients. (g)$\sim$(i) is the accuracy of 8 clients.}
			\label{fig:acc_fashionmnist}
		\end{figure*}

		Table \ref{table:pqfl_acc_personal_fashionmnist} demonstrates the average model accuracy of clients in PQFL is higher than that without personalized layer in the most settings. Further, we can see that when $\alpha$ is constant, the average accuracy improves as the clients quantity increases. Fig. \ref{fig:acc_fashionmnist} shows the accuracy of client models with personalization layer is generally higher on the testset compared to those without personalization layer. This indicates that the client models are more personalized than the global model based on the experimental results.
		
		\begin{table}[htbp]
			\begin{center}  
				\caption{Average accuracy of client model with and without personalized layer.}
				\label{table:pqfl_acc_personal_fashionmnist}
				\setlength{\tabcolsep}{2mm}
				\renewcommand{\arraystretch}{1}
				\begin{threeparttable}
					\begin{tabular}{ c  c  c  c}
						\toprule 
						$\alpha$ & $M$ & {With personalized layer} & {Without personalized layer}\\
						\midrule
						\multirow{3}{*}{1} &2&\textbf{97.0\%}&94.1\%\\
						&4&\textbf{97.1\%}&95.9\%	\\
						&8&\textbf{97.7\%}&96.1\%	\\
						\midrule
						\multirow{3}{*}{10} &2&\textbf{94.5\%}&94.5\%\\
						&4&\textbf{96.2\%}&96.1\%	\\
						&8&\textbf{97.2\%}&96.8\%	\\
						\midrule
						\multirow{3}{*}{100} &2&\textbf{96.5\%}&94.5\%\\
						&4&\textbf{96.8\%}&96.0\%	\\
						&8&\textbf{97.0\%}&96.8\%	\\
						\bottomrule
					\end{tabular}
					\begin{tablenotes}
					\footnotesize
					\item $M$ is the number of clients.
					\end{tablenotes}
			\end{threeparttable}
			\end{center}
		\end{table}

	\subsection{Convergence Analysis}	
\begin{figure*}[htbp]
			\centering
			\includegraphics[width=0.3\textwidth]{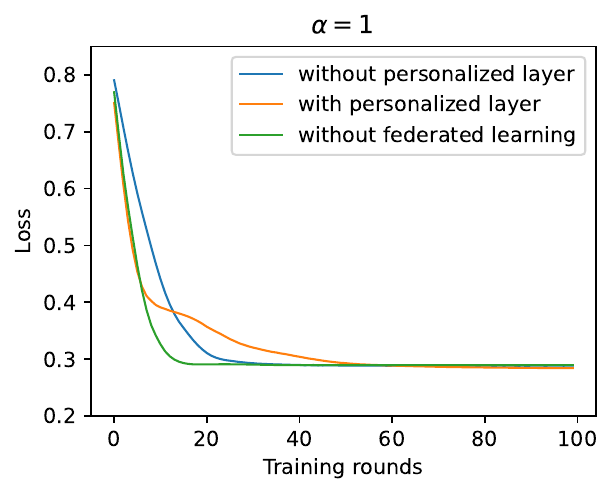}
			\includegraphics[width=0.3\textwidth]{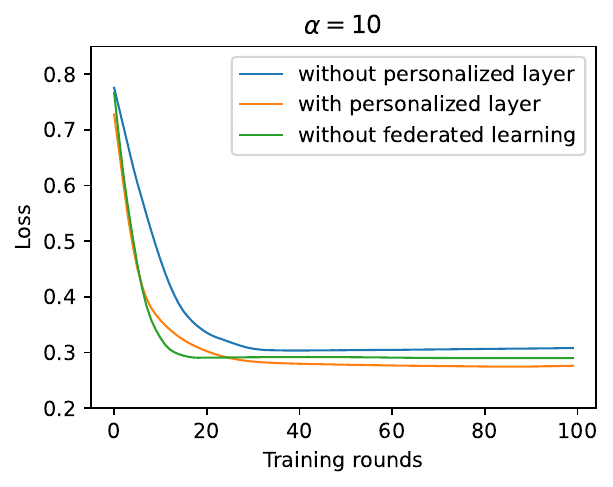}
			\includegraphics[width=0.3\textwidth]{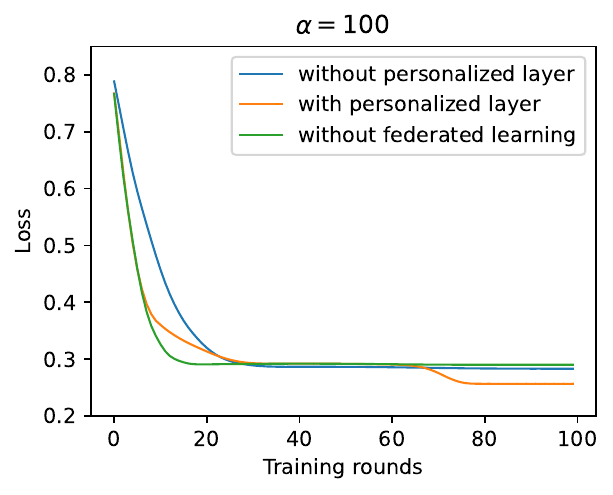}\\
			(a)\qquad\qquad\qquad\qquad\qquad\qquad\qquad(b)\qquad\qquad\qquad\qquad\qquad\qquad\qquad(c)\\
			\includegraphics[width=0.3\textwidth]{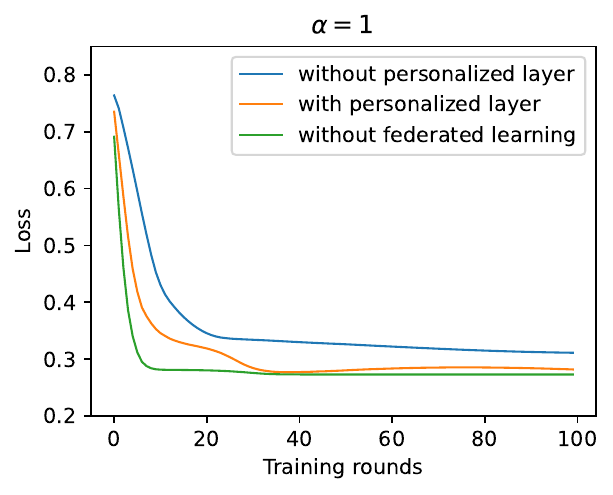}
			\includegraphics[width=0.3\textwidth]{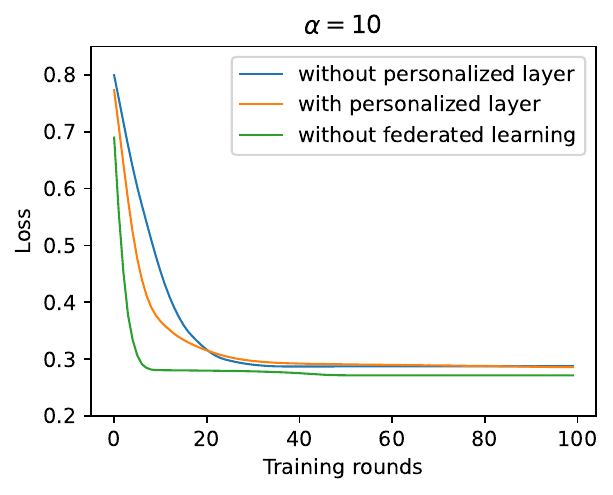}
			\includegraphics[width=0.3\textwidth]{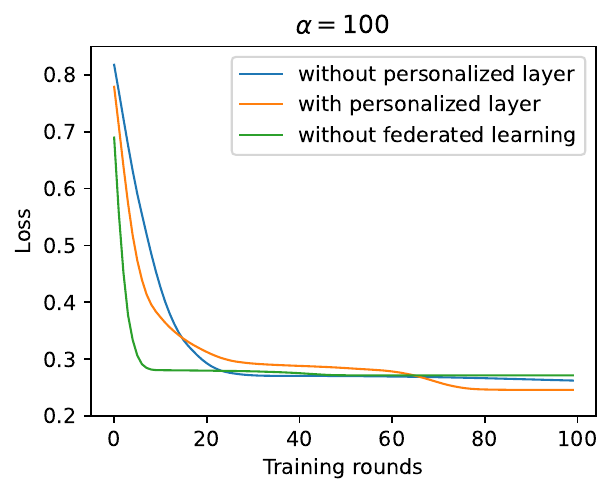}\\
			(d)\qquad\qquad\qquad\qquad\qquad\qquad\qquad(e)\qquad\qquad\qquad\qquad\qquad\qquad\qquad(f)\\
			\includegraphics[width=0.3\textwidth]{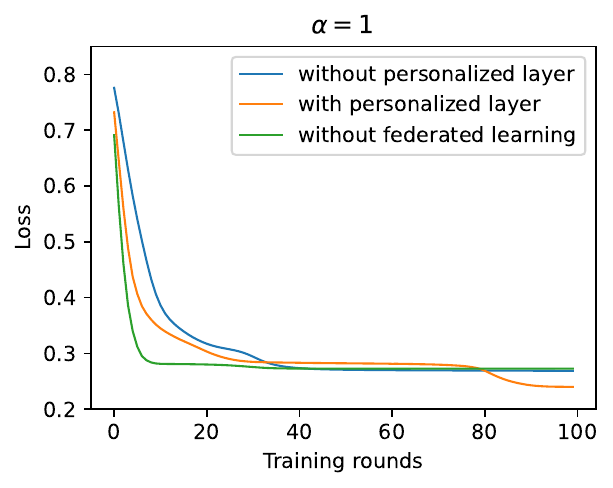}
			\includegraphics[width=0.3\textwidth]{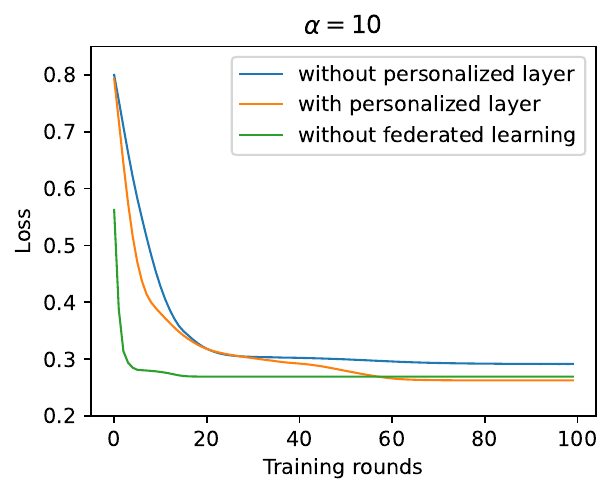}
			\includegraphics[width=0.3\textwidth]{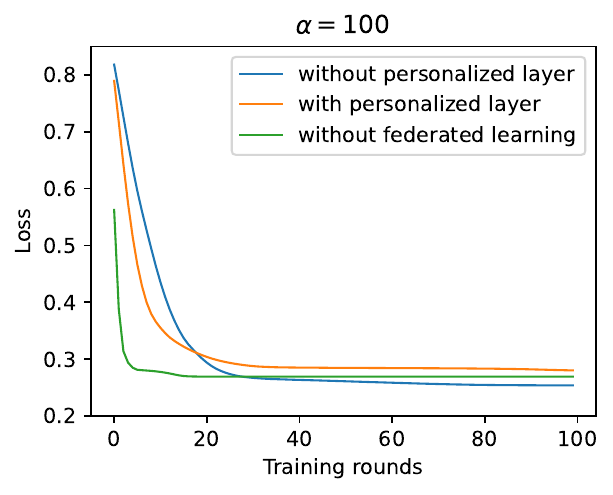}\\
			(g)\qquad\qquad\qquad\qquad\qquad\qquad\qquad(h)\qquad\qquad\qquad\qquad\qquad\qquad\qquad(i)
			\caption{Loss of the server with and without personalized layer on FashionMNIST dataset when $\alpha=1,10,100$, respectively. (a)$\sim$(c) is the loss of 2 clients. (d)$\sim$(f) is the loss of 4 clients. (g)$\sim$(i) is the loss of 8 clients.}
			\label{fig:loss_fashionmnist}
		\end{figure*}

		We carry out experiments involving 2, 4, and 8 clients in the training process. By minimizing the loss function, the model parameters stabilize and achieve optimal performance, leading to model convergence. Fig. \ref{fig:loss_fashionmnist} illustrates the convergence speed of the loss function for the following three scenarios to assess the convergence of the PQFL algorithm:
		
		(1) With personalized layer. The proposed PQFL algorithm includes a client model with both the base layer and a personalized layer, while the server only contains the base layer.
		
		(2) Without personalized layer. A quantum federated learning where the server and client models only include the base layer and do not include the personalized layer.
		 
		(3) Without federated learning. Training on a quantum neural network with only the base layer.
		
		It suggests that the PQFL algorithm successfully converges while consistently achieving comparable performance in the other two scenarios.
	\subsection{Overhead Analysis}	  
		The overhead of the PQFL algorithm consists of both communication and computation overhead. Table \ref{table:pqfl_communication} indicates the communication overhead of the PQFL algorithm, which can be analyzed through storage overhead and time overhead. For the server, $4nk$ base layer parameters need to be encoded into the quantum state and sent to the $M$ clients. The total qubits sent from sever to $M$ clients in $N$ global training rounds is $4nkNM$. Suppose performing an encoding or decoding operator costs $t_c(\sim 2.5\times10^{-8}s)$ \cite{Frank2019}, and transmitting a qubit from the server to the client through the quantum channel costs $t_n(\sim 10^{-3}s)$ \cite{Simon2007}, the downlink time of training $N$ rounds is:
		\begin{equation}\label{Eq:time_downlink}
			\begin{aligned}
				T_{down}&\approx 4nkNt_c + 4nkNMt_c + Nt_n\\
				&= 4nkNt_c(M+1)+Nt_n.
			\end{aligned}
		\end{equation}
		For a client, there are $4nk$ base layer parameters that need to be encoded into the quantum state and sent to the server. The total qubits sent from a client in $N$ global training rounds is $4nkN$. The uplink time of training $N$ rounds is:
		\begin{equation}\label{Eq:time_uplink}
			T_{up}\approx 4nkN\cdot 2t_c + Nt_n = 8nkNt_c+Nt_n.
		\end{equation}
		\begin{table}[htbp]
			\begin{center}
				\caption{Communication overhead of server and client.}
				\label{table:pqfl_communication}
				\setlength{\tabcolsep}{3mm}
				\begin{tabular}{c  c  c }
					\toprule
					&   {Server}& {Client}\\  
					\midrule
					{Storage(qubit)} &  {$4nkNM$} & {$4nkN$}\\
					{Time(s)} &  {$4nkNt_c(M+1)+Nt_n$} & {$8nkNt_c+Nt_n$}\\
					\bottomrule
				\end{tabular}
			\end{center}
		\end{table}
		
		The computation overhead of the server and client is shown in Table \ref{table:pqfl_computation}. For an $L\times L$ image, the server and the clients need $2\log_2L$ qubits to encode an image into quantum state. The circuit depth of the server and clients depends on how many layers the base layer has. The server model's depth is $k$ when the base layer is repeated $k$ times, and the depth of the client model is $k+1$ for the $k$ base layers and one personalized layer.
		\begin{table}[htbp]
			\begin{center}
				\caption{Computation overhead of server and client.}
				\label{table:pqfl_computation}
				\setlength{\tabcolsep}{8mm}
				\begin{tabular}{c  c  c }
					\toprule
					& {Server}& {Client}\\  
					\midrule
					{Circuit depth} &  {$k$} & {$k+1$}\\ 
					{Circuit qubit} &  {$2\log_2L$} & {$2\log_2L$}\\ 
					\bottomrule
				\end{tabular}
			\end{center}
		\end{table}	
		
		Parameters of server and client models are encoded into GHZ states instead of encrypted by homomorphic encryption which demands a lot of computation. The number of entangled qubits of GHZ state is related to the number of model parameters, which determine the time cost and accuracy of model parameters measuring. Accurate inference of the model parameter values encoded in the quantum state can only be achieved when the corresponding estimator is highly precise, such as when the variance is minimal. The relationship of the measurement times $R$ and variance $\sigma^2$ of the measurement results can be expressed as follows.
		\begin{equation}\label{Eq:variance}
			\sigma^2 \sim\frac{1}{R}.
		\end{equation}
		The variance declines with the increase of $R$, and the precision of the model parameter estimation is increased, too.

	\subsection{Security Analysis}
		The participants in this scenario are semi-honest, that is to say, the client involved in training and the central server responsible for aggregation will honestly follow the communication protocol, but they will both try to extract as much information as possible from the data transmitted during the interaction.
		
		\textbf{Theorem 1.} \textit{External attacks resistance. The external adversary can not obtain the model parameters and privacy images of clients.}
		
		\textit{Proof.} If the external adversary wants to obtain  private information about the server and clients, he has to copy quantum states, otherwise, the model parameters encoded in quantum states can not be obtained by the adversary. The quantum non-cloning theorem states that it is infeasible to exactly duplicate an unknown quantum state, which means an external attacker can not copy a quantum state without being discovered by the server and clients. Suppose an external adversary can copy a quantum state $|\varphi\rangle$ by performing a unitary $U_{copy}$ on $|0\rangle$, which can be denoted as:
		\begin{equation}\label{Eq:copy1}
			U_{copy}|\varphi\rangle|0\rangle=|\varphi\rangle|\varphi\rangle.
		\end{equation}
		Given two quantum states $|\psi_1\rangle$ and $|\psi_2\rangle$, it can be obtained by:
		\begin{equation}\label{Eq:copy2}
			\begin{aligned}
				U_{copy}|\psi_1\rangle|0\rangle=|\psi_1\rangle|\psi_1\rangle,\\
				U_{copy}|\psi_2\rangle|0\rangle=|\psi_2\rangle|\psi_2\rangle.		
			\end{aligned}
		\end{equation}
		If the external adversary wants to copy an arbitrary quantum superposition state $|\varphi\rangle=c_1|\psi_1\rangle+c_2|\psi_2\rangle$, where $|c_1|^2+|c_2|^2=1$, using the operator $U_{copy}$, the expected result should be:
		\begin{equation}\label{Eq:copy3}
			\begin{aligned}
				U_{copy}|\varphi\rangle|0\rangle=&|\varphi\rangle|\varphi\rangle\\
										 =&(c_1|\psi_1\rangle+c_2|\psi_2\rangle)(c_1|\psi_1\rangle+c_2|\psi_2\rangle)\\
										 =&c_1^2|\psi_1\rangle|\psi_1\rangle+c_1c_2|\psi_1\rangle|\psi_2\rangle+\\
										 &c_2c_1|\psi_2\rangle|\psi_1\rangle+c_2^2|\psi_2\rangle|\psi_2\rangle.
			\end{aligned} 
		\end{equation}
		However, the result can be also calculated as follows:
		\begin{equation}\label{Eq:copy4}
			\begin{aligned}
				U_{copy}|\varphi\rangle|0\rangle&=U_{copy}(c_1|\psi_1\rangle+c_2 |\psi_2\rangle)|0\rangle\\
										 &=(c_1 U_{copy}|\psi_1\rangle|0\rangle+c_2 U_{copy}|\psi_2\rangle)|0\rangle\\
										 &=c_1|\psi_1\rangle|c_1\rangle+c_2|\psi_2\rangle|\psi_2\rangle.
			\end{aligned} 
		\end{equation}
		The results from Eq. (\ref{Eq:copy3}) and Eq. (\ref{Eq:copy4}) are conflicting, so the unitary operator $U_{copy}$ that can copy any quantum state does not exist. This means that an external attacker can not copy the quantum state transmitted between the server and the client, so the model parameters cannot be measured. In addition, the clients' privacy images are only retained locally and are not transmitted, so external attacks can not reverse the privacy images through the model parameters, too.
		
		\textbf{Theorem 2.} \textit{Internal attacks resistance. The server and clients can not obtain any privacy information about each other.} 
		
		\textit{Proof.} The initiator of the internal attack is the server and the clients. On the one hand, the server can obtain the quantum state $|\varPsi_{client}\rangle$ which encoded the client model parameters.
		\begin{equation}\label{Eq:avg_para}
			\begin{aligned}
				|\varPsi_{client}\rangle &= \otimes_{i=1}^K|\varPsi_i\rangle \\
				&= \frac{1}{\sqrt{2}^K}\otimes_{i=1}^K(|00\cdots0\rangle+e^{i\sum_{m=1}^{M}(F_m\cdot\theta_{m,i})}|11\cdots1\rangle),
			\end{aligned} 
		\end{equation}
		where $F_m$ is the weighted score of the $m$th client, $M$ denotes the clients quantity, and $K$ represents the quantity of the base layer parameters. Thus the server can only get the weighted average parameters $\bar{\boldsymbol{\theta}}=\{\bar\theta_1,...,\bar\theta_K\}$ of the base layer, where $\bar{\theta_i} = \sum_{m=1}^{M}(F_m\cdot\theta_{m,i})$, but it is not clear about the personalized layer parameters of each client. 
		
		On the other hand, the clients can obtain nothing from the GHZ state that is encoded by all clients, because the quantum state can be expressed as:
		\begin{equation}\label{Eq:client_identity}
			\begin{aligned}
				\rho =& |\varPsi_{client}\rangle\langle\varPsi_{client}| \\
				=& \otimes_{i=1}^K|\varPsi_i\rangle\langle\varPsi_i| \\
				= &\frac{1}{2^K}\otimes_{i=0}^K(|00\cdots0\rangle+e^{i\sum_{m=1}^{M}(F_m\cdot\theta_{m,i})}|11\cdots1\rangle)\\
				&(\langle00\cdots0|+e^{i\sum_{m=1}^{M}(F_m\cdot\theta_{m,i})}\langle0011\cdots1|)\\
				=&\frac{\mathbb{I}}{2},
			\end{aligned} 
		\end{equation}
		which is a completely mixed state. This means that each qubit from the encoded GHZ state reveals nothing about each client’s privacy information.
		Therefore, the proposed PQFL algorithm has high security which can resist both external attacks and internal attacks.
	
	\subsection{Comparison}	
		The security comparison between this scheme and other representative quantum federated learning schemes is shown in Table \ref{table:pqfl_secure_compare}. 
		
		\textbf{VQA-QFL} \cite{Huang2022} performs quantum federated learning on a non-IID dataset, but does not adopt privacy protection technology to ensure the security of the model.		
		
		\textbf{DTQFL} \cite{10210652} is a quantum federated learning model with additional personalized training on the local model after the completion of global training, the same does not protect the gradient information transmitted between the client and the server, which can not resist internal and external attacks. 
		
		\textbf{QFL-BQC} \cite{Li2021} is a federated learning model based on differential privacy, where noise needs to be introduced to ensure its security at the expense of model accuracy. It resides gradient attacks by adding noise and gradient tailoring to the gradient, which will sacrifice the model accuracy, and does not take into account the personalization of the client model. 
				
		\textbf{CryptoQFL} \cite{10313628} is an FL model based on classical homomorphic encryption, additional keys need to be generated to encrypt the gradients or parameters of the model, increasing the computational complexity and communication overhead to ensure its security.
		\begin{table}[htbp]
			\caption{Comparison with other quantum federation learning schemes.}
			\label{table:pqfl_secure_compare}
			\centering
			\begin{tabular}{ccc}
				\toprule
				{Scheme}&{Privacy preserving}& {Personalization} \\
				\midrule 
				{VQA-QFL \cite{Huang2022}}&\usym{2717}&\usym{2717} \\
				{DTQFL \cite{10210652}}&\usym{2717}&\Checkmark  \\
				{QFL-BQC \cite{Li2021}}&\Checkmark&\usym{2717}\\
				{CryptoQFL \cite{10313628}}&\Checkmark&\usym{2717} \\
				{Ours}&\Checkmark&\Checkmark \\
				\bottomrule
			\end{tabular}
		\end{table}
		Our scheme uses the quantum parameter weighted average algorithm to ensure transmission parameters are not leaked through the quantum channel, and can simultaneously resist internal and external attacks. In the training process, the client model is personalized without additional local training.
		Thus, our proposed privacy image classification scheme based on the personalized quantum federated learning algorithm has high classification accuracy and strong security. Additionally, the client model can also meet personalized needs even with an unbalanced dataset distribution.

\section{Conclusion}\label{section_v}
	To address the issue of lack of personalization on the clients, this paper presents a PQFL algorithm, and applies it to privacy image classification. The personalized layer of the QNN model of the clients can help the client obtain a personalized model even with the non-IID data. The quantum parameter weighted average aggregation algorithm secures both the global and local models as well as the privacy data. 
	The experimental results demonstrate that the proposed PQFL algorithm is capable of  classifying the privacy images effectively and make the client model more personalized, ensuring the data privacy and security at the same time. With the continuous development of quantum hardware, the proposed personalized quantum federated learning model promotes the further development of distributed quantum machine learning and holds potential for broader applications in fields, ranging from healthcare and finance to communication and cybersecurity, ultimately shaping the landscape of quantum artificial intelligent systems.
	

\bibliographystyle{IEEEtran}
\bibliography{ref}

\vfill

\end{document}